\documentclass[a4paper]{article}

\usepackage{INTERSPEECH2022,multirow,booktabs,enumerate,siunitx,hyperref,cleveref}
\usepackage[inline]{enumitem}

\title{Understanding temporally weakly supervised training: A case study for keyword spotting}
\name{Heinrich Dinkel, Weiji Zhuang, Zhiyong Yan, Yongqing Wang, Junbo Zhang and Yujun Wang}
\address{
  Xiaomi Corperation, Beijing, China}
\email{\{dinkelheinrich,zhuangweiji,yanzhiyong,wangyongqing3,zhangjunbo1,wangyujun\}@xiaomi.com}

\begin{document}

\maketitle
\begin{abstract}
The currently most prominent algorithm to train keyword spotting (KWS) models with deep neural networks (DNNs) requires strong supervision i.e., precise knowledge of the spoken keyword location in time.
Thus, most KWS approaches treat the presence of redundant data, such as noise, within their training set as an obstacle.
A common training paradigm to deal with data redundancies is to use temporally weakly supervised learning, which only requires providing labels on a coarse scale.
This study explores the limits of DNN training using temporally weak labeling with applications in KWS.
We train a simple end-to-end classifier on the common Google Speech Commands dataset with increased difficulty by randomly appending and adding noise to the training dataset.
Our results indicate that temporally weak labeling can achieve comparable results to strongly supervised baselines while having a less stringent labeling requirement.
In the presence of noise, weakly supervised models are capable to localize and extract target keywords without explicit supervision, leading to a performance increase compared to strongly supervised approaches.
\end{abstract}

\section{Introduction}

Keyword spotting (KWS) is nowadays a crucial task for the deployment of intelligent voice assistants and acts as the gateway between a user's request and back-end services in the cloud.
Most methods for KWS require strong supervision i.e., the exact start and end timesteps of a spoken utterance for training, which is mainly acquired by a forced alignment using an automatic speech recognition (ASR) system~\cite{chen2014small,sun2017compressed,majumdar2020matchboxnet}.
However, forced alignment may fail if the audio is heavily noise degraded, leading to misaligned labels, which can limit detection accuracy~\cite{park2020learning,lopatka2020state,chen2018sequence}.
In those cases where forced alignment is difficult, one possible solution is to manually label the presence of a keyword with precise on and offsets.
Still, precise manual labeling is a costly and laborious effort.
Another, less costly alternative is weak labeling, which only requires labeling the presence of a keyword within a coarse time-range e.g., within 5 seconds.
% In other words, large parts of the input data presented to the model are irrelevant i.e., noise, to the task.
This work is an initial study on the effects of temporally weak labeling for end-to-end (E2E) KWS.
Weakly supervised learning has attracted success in other audio-related fields, such as audio tagging~\cite{kong2020panns,cances_icassp22_semisupervised_at,gong2021psla}, voice activity detection~\cite{chen2020voice,dinkel2021voice,xu2021lightweight} and sound event detection~\cite{dinkel2021towards}.
Previous work of max-pooling loss or sequence-to-sequence training~\cite{sun2016max,an2019robust,wang2020wake,jose2020accurate,wang2021wake} treat imperfect label boundaries as a major hindrance to the performance of a KWS system.

% This work is motivated to facilitate the training of real-world deployable keyword spotting models.
In our point of view, training KWS models with strong supervision restricts their generalization, since the model is explicitly told what the target label is and when the label is present, making training trivial.
We argue, that, training a model in temporally weakly supervised fashion requires the model not only to discover the target label, but also to localize it.
We, therefore, seek to answer the following questions in this work: 
\begin{enumerate*}[label=\Roman{*}.]
  \item Are strongly supervised KWS models superior to weakly supervised ones?
  \item Can a neural network automatically implicitly learn the presence of a target label, even when hidden in noise?
  \item Is the removal of noise/silence important for KWS training?
\end{enumerate*}
Our contributions are:
\begin{itemize}
    \item A study of the impact of temporally weakly supervised training for KWS.
    \item A comparison between strongly supervised KWS and temporally weakly supervised KWS.
    \item A set of suggestions to successfully train temporally weakly supervised KWS models.
\end{itemize}
This paper is structured as follows.
\Cref{sec:framework} introduces this works training framework. 
Then the experimental setup is shown in \Cref{sec:experiments} and the results are displayed in \Cref{sec:results}.
A brief discussion regarding our results is seen in \Cref{sec:discussion} and the work is concluded in \Cref{sec:conclusion}.

\section{Framework}
\label{sec:framework}

Our training framework aims to investigate the effects of weak labeling and consequently weakly supervised training on KWS models.
Specifically, given the keyword label set $\mathcal{K} = \{0,\ldots, K-1\}$ with $K$ keywords, our study focuses on the impact of different label qualities provided to the model at different scales.
We define strong labels in the context of KWS as frame-level targets, such that for each input frame $x_i \mapsto y_i$ for an input audio spectrogram $x_{1:T}$ with $T$ frames and $y_i \in \mathcal{K}$ is the corresponding label.
In other words, strong labels know the exact location in time of a target label.
On the contrary, weak labels are defined on a coarse scale (e.g., for each audio clip), meaning that only a single label corresponds to an audio clip as $x_{1:T} \mapsto y \in \mathcal{K}$.
In a sense, common E2E approaches in KWS are weakly-supervised, since only a single target label is provided for an input sequence.
However, preprocessing for E2E training generally removes undesirable data, such as silence or other noises, such that the input audio sample only contains the target keyword.
\begin{figure*}[tb]
    \centering
    \includegraphics[width=0.85\linewidth]{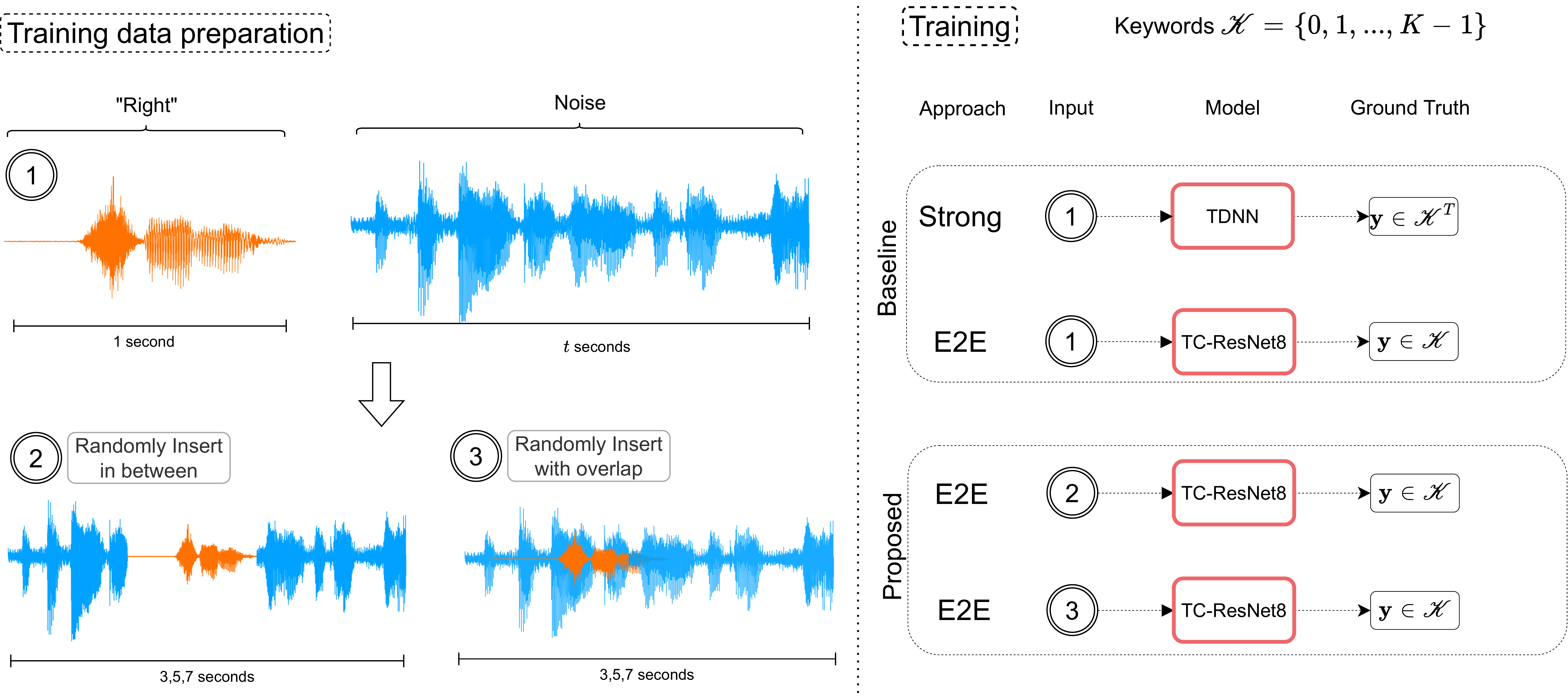}
    \caption{The framework in this work focuses on using three distinct datasets: First, the original clean keyword spotting dataset. Second, we randomly insert keywords between noisy samples without overlap. Third, insertion with overlap using a specified signal-to-noise (SNR) ratio. Best viewed in color.}
    \label{fig:framework}
\end{figure*}
To facilitate temporally weakly-supervised research, we use three different training datasets.
\begin{enumerate*}
    \item A clean KWS dataset as the baseline approach.
    \item A temporally weak version of the aforementioned dataset, where the target keyword is randomly injected into a noise sequence of a specified length without overlap.
    \item Additionally overlapping the noise-sequence with the target keyword.
\end{enumerate*}
% Then, we train a strongly-labeled baseline and our proposed end-to-end approach.
Our framework can be seen in \Cref{fig:framework}.
Note that different from traditional approaches that our experimental setup uses real-world noises, which increases the difficulty for training significantly.
% The noise dataset in this work is chosen to be Audioset~\cite{gemmeke2017audio}.
% Please note that we randomly temporally append noise for data generation, which could also include spoken utterances containing target keywords, increasing the training difficulty.

\section{Experiments}
\label{sec:experiments}

\subsection{Dataset and data preparation}
\label{ssec:dataset_prepare}

This work mainly uses the Google Speech Commands V1 (GSCV1)~\cite{warden2018speech} dataset as the KWS dataset of choice.
GSCV1 contains 65,000 utterances of 30 different keywords, spoken by thousands of people.
We use the common 11 class subset of GSCV1, where the original 30 classes have been reduced to 10 common keywords: ``Yes'', ``No'', ``Up'', ``Down'', ``Left'', ``Right'', ``On'', ``Off'', ``Stop'', ``Go'' while the rest 20 keywords are mapped to ``unknown/noise''.
The official training/validation/testing split containing 51,088/6,798/6,835 utterances, respectively, is used in this study.
Please note that all experiments are evaluated on the clean test set of GSCV1 (6,835 utterances) and only the training subset is modified.

To simulate real-world noises, we use Audioset~\cite{gemmeke2017audio} as our main external noise dataset.
Audioset contains $\approx$ 5200 hours of unconstrained audio data, where most samples contain speech and/or music.
We randomly sample 51,088 audio clips from Audioset to modify the training dataset, as it is explained in \Cref{ssec:dataset_prep}.

\subsubsection{Dataset preparation}
\label{ssec:dataset_prep}

To study the effects of weak labeling, we modify the original GSCV1 training dataset.
Overall, three training datasets are used in this work.
First, the original GSCV1 dataset, consisting of clean at most 1 second long utterances.
Note that we did not inject additional noise provided in the dataset into the training data.
Second, we generate our weakly labeled dataset denoted as ``Weak-$t$s'', where $t = 3,5,7$. This dataset is created by firstly uniformly sampling an audio clip from Audioset.
Then, we randomly sample the insertion point which is drawn from $\mathbf{U}(0,t-L)$, where $t$ represents the target sample length i.e., 3 seconds, $L$ the keyword length ($\leq 1 $s) and $\mathbf{U}$ is the uniform distribution.
Third, we generate the dataset $\text{Weak}_{\text{SNR}}\text{-}t$s by mixing a keyword sample with an audio clip by a given signal-to-noise (SNR) ratio, using the same creation pipeline as for Weak-$t$s.
We overall experiment with three fixed SNRs ($0,5,10$ dB) and therefore create three subsets.
% It is important to state that while our work adds additional noise data 
It is important to state that our work only enlarges each sample's duration, but keeps the amount of labels identical.

\subsection{Setup}

\subsubsection{Strongly labeled baseline}

We use the state-of-the-art regular LF-MMI approach as proposed in~\cite{wang2020wake} with a time-delay neural network (TDNN) as our baseline approach.
Specifically, our implementation follows the Kaldi~\cite{povey2011kaldi} toolkit's KWS recipes as described in~\cite{wang2020wake}.
To provide comparable results between strongly supervised methods and E2E methods, we device two settings: oracle and force-align.
Oracle uses the precise on- and offset timestamps generated during data preparation, while the force-align (FA) setting predicts the location of the keyword within an audio sequence.
The oracle setting symbolizes having access to a perfect alignment, while the FA setting provides insights into the effect of flawed alignments in KWS.

\subsubsection{End-to-end approaches}

For all E2E experiments, we utilize the state-of-the-art TC-ResNet8~\cite{Choi2019} model as our back-end, due to its small parameter size of 66,000 and its fast training time.
The small parameter size also means that our experiments are unlikely a result of the model being able to overfit the training data, which is often the case for larger models.

Regarding front-end feature extraction, all audio clips are converted to a 16 kHz sampling rate. 
We use log-Mel spectrograms (LMS) with 64 bins extracted every 10 ms with a window of 32ms.
Batch-wise zero padding to the longest clip within a batch is applied for all experiments.
Training runs with a batch size of 64 for at most 200 epochs using Adam optimization~\cite{kingma2014adam} with a learning rate of 0.001.
We choose the standard categorical cross-entropy loss as our training objective.
The top-4 model checkpoints achieving the highest accuracy on the held-out validation dataset are weight-averaged and evaluated on the testing set.
The neural network back-end is implemented in Pytorch~\cite{PaszkePytorch}.

\subsubsection{Metrics}

To compare our results with other works on this dataset, we use accuracy as the main evaluation metric in this study.
Further, macro (class) averaged mean average precision (mAP) is used as a secondary measure to represent a class-independent measure.
Note that mAP is only provided for our proposed E2E models, since the traditional HMM-based baseline does not output a score for each segment.

\section{Results}
\label{sec:results}

\subsection{Effects of weakly supervised training}
\label{ssec:weakly_supervised_training}

First, we study the effects of weakly supervised training and compare our results to strongly-supervised baselines.
The results are provided in \Cref{tab:weakly_supervised_baseline}.
\begin{table}[htbp]
    \centering
    \begin{tabular}{ll|rrrr}
    \toprule
    Approach & Data & Accuracy & mAP \\
        \midrule
    Oracle & \multirow{3}{*}{Clean} & 96.23 & - \\
    FA &  & 96.21 &  - \\
    Ours &  & \textbf{97.03} & \textbf{98.28} \\
        \hline
    % Oracle &\multirow{3}{*}{Clean}
    % Strong & 
    Oracle & \multirow{3}{*}{Weak-3s} & 95.39 & -\\
    FA &  & 95.08  & - \\
    Ours & & {96.36} & 96.93\\
    \hline
    Oracle & \multirow{3}{*}{Weak-5s} & 96.86 &- \\
    FA & - & 93.43  & - \\
    Ours &  & {96.74} & 96.87 \\
    \hline
    Oracle & \multirow{3}{*}{Weak-7s} & 96.13 & - \\
    FA &  & 87.21  & - \\
    Ours &  & {96.68} & 96.27 \\
        \bottomrule
    \end{tabular}
    \caption{Results using our our proposed weakly-supervised training regime using different input-sample lengths. Evaluation is done exclusively on the clean testset of GSCV1. ``-'' represents not available. Best results are highlighted in bold.}
    \label{tab:weakly_supervised_baseline}
\end{table}
First, it can be seen that our baseline approach (Oracle/FA) achieves a near-identical performance (96.23\% and 96.21\%) when trained with clean data.
Furthermore, when appending noise to the clean keyword set (3,5,7s), performance consistently decreases for our strong baseline, falling from 96.21\% to 87.21\%.
However, unsurprisingly, the oracle approach's performance remains relatively stable.
These results indicate the difficulty of using FA in real-world scenarios.

On the contrary, our proposed E2E approach consistently performs well across all tested scenarios.
Increasing the input audio length from 3 s to 7s only marginally decreases performance from 97.03 \% to 96.68\%.
However, the mAP declines from 98.28\% to 96.27\% between the clean baseline and the 7s train setting.
We conclude that a weakly supervised model is capable of automatically localizing the target keyword without explicit supervision.

\subsection{Train and test duration mismatch}
\label{ssec:train_test_crops}

The previous results in \Cref{tab:weakly_supervised_baseline} show that weakly-supervised training is indeed capable of achieving similar accuracies as to their strongly supervised counter-parts.
However, in terms of mAP, a significant performance drop has been observed.
We believe that the core reason for this behavior is the mismatch between training and evaluation dataset durations, i.e., weakly-supervised models are trained on 3,5,7 s of audio data, but evaluated on 1 s long samples.
Thus, we further evaluate whether random cropping of the training samples is advantageous in this regard.

\begin{table}[htbp]
    \centering
    \begin{tabular}{ll|rrrr}
    \toprule
    Approach & Data & Crop & Accuracy & mAP \\
        \midrule
        Oracle & \multirow{3}{*}{Clean} &  & 96.23 & - \\
        FA &  &  & 96.21  & -  \\
        Ours &  &  & 97.03 & 98.28 \\
        \hline
         \multirow{3}{*}{Ours} & Weak-3s & \multirow{3}{*}{1s} & \bf{97.25} & \bf{98.58}\\
         & Weak-5s & &  96.75 & 98.18 \\
         & Weak-7s &  & 97.10 & 98.54 \\
        \bottomrule
    \end{tabular}
    \caption{Experiments using random crops during training compared to our baselines. ``-'' represents not available. Best results in bold.}
    \label{tab:train_test_crops_mismatch}
\end{table}

The results when utilizing random cropping are provided in \Cref{tab:train_test_crops_mismatch}.
It can be observed that random cropping enhances performance across all tested weakly supervised E2E approaches.
There are two main reasons for this behavior: First and foremost, when using random cropping, the training and testing audio clip duration match, which yields higher confidence for each evaluation sample, evident at the increased mAP scores.
Second, random cropping increases the training sample size, which can also lead to a performance improvement.
We conclude that weakly labeled E2E models can achieve similar performance to their strongly supervised counterparts.
\begin{figure}[tb]
    \centering
    \includegraphics[width=0.90\linewidth]{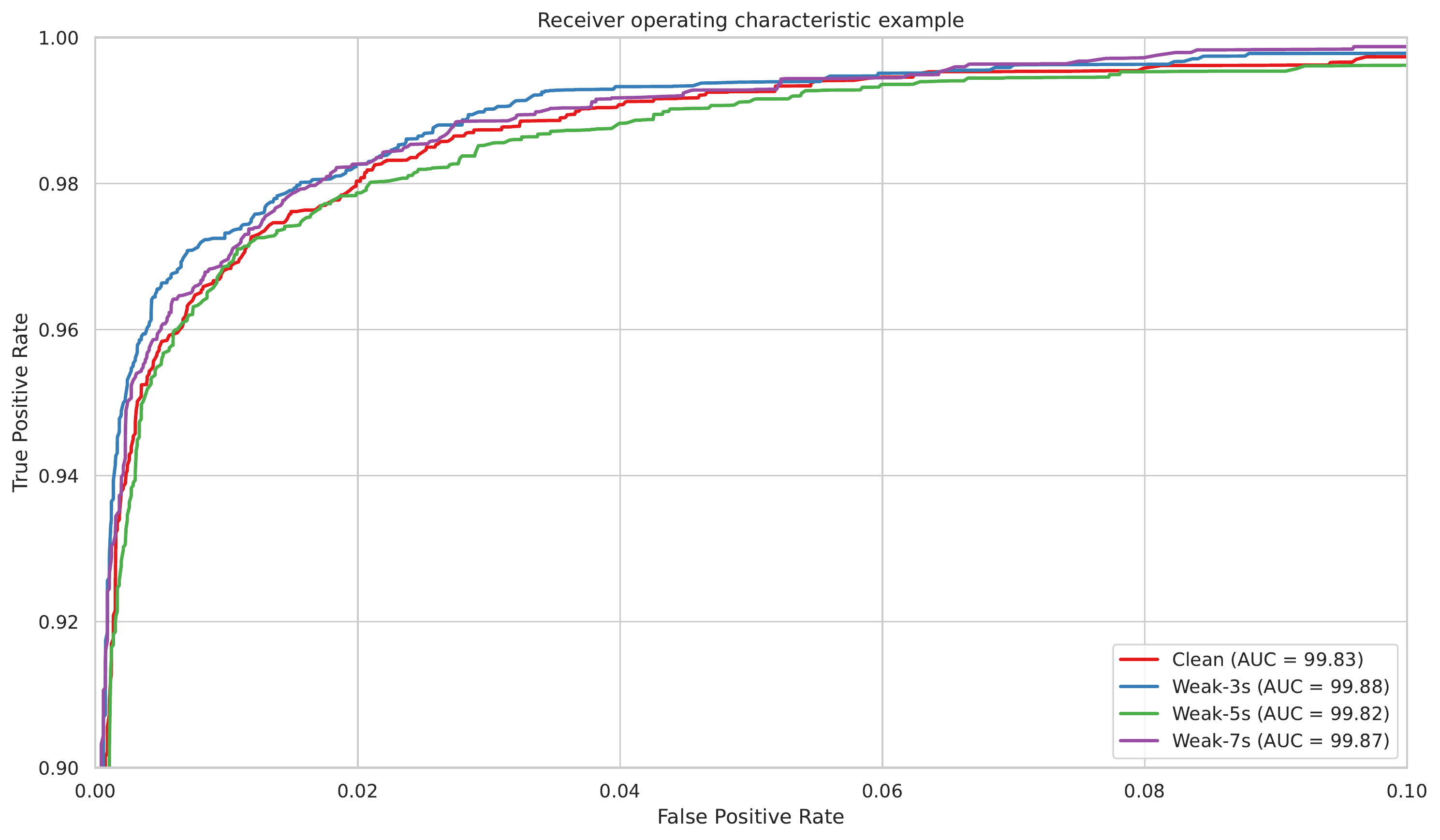}
    \caption{ROC curves for our proposed E2E models with corresponding AUC values. Best viewed in color.}
    \label{fig:roc_curves}
\end{figure}
Finally, we also provide the receiver operating characteristic (ROC) curves in \Cref{fig:roc_curves}.

\subsection{Additive noise}
\label{ssec:additive_noise}

A core problem with the previous experiments in \Cref{ssec:train_test_crops} is that the simulated training datasets are purely fabricated, where the clean target keyword is present within a noisy audio clip.
This leads to the question of whether the model is in fact capable of localizing a keyword or if the model overfits to the artificially clean target keyword segment.
Thus, we investigate whether the model can localize a target keyword hidden within the noise. 
Here, we use the $\text{Weak}_{\text{SNR}}-t$s datasets for training.

\begin{table}[htbp]
    \centering
    \begin{tabular}{ll|rrr}
    \toprule
    Data & SNR & Oracle & FA & Ours\\
        \midrule
        Clean & $\infty$ & 96.23 & 96.21 & \textbf{97.03}\\
        \hline
        $\text{Weak}_{\text{SNR}}$-3s & \multirow{3}{*}{0} &  92.90 & 81.65 & \textbf{93.69}\\
        $\text{Weak}_{\text{SNR}}$-5s &  & 90.40  & 73.06 & \textbf{90.78}\\
        $\text{Weak}_{\text{SNR}}$-7s & & \textbf{85.98} & {67.29} & 62.44 \\
        \hline
        $\text{Weak}_{\text{SNR}}$-3s & \multirow{3}{*}{5} & 94.20 & 89.37 & \textbf{95.25}\\
        $\text{Weak}_{\text{SNR}}$-5s & & 94.95 & 84.48 & \textbf{94.06} \\
        $\text{Weak}_{\text{SNR}}$-7s & & 88.61 & 72.59 & \textbf{93.45} \\
        \hline        
        $\text{Weak}_{\text{SNR}}$-3s & \multirow{3}{*}{10} & 95.42 & 92.51  & \textbf{96.36}\\
        $\text{Weak}_{\text{SNR}}$-5s & & 94.74 & 87.37 & \textbf{95.29}\\

        $\text{Weak}_{\text{SNR}}$-7s & & \textbf{94.66} & 78.17 & {94.56} \\
        \bottomrule
    \end{tabular}
    \caption{Experiments using keywords being entirely overlapped with noise given a specified SNR. The ``Ours'' approach uses 1s random crops. Best in bold.}
    \label{tab:noisy_experiments}
\end{table}

% \begin{table}[htbp]
%     \centering
%     \begin{tabular}{ll|rrr}
%     \toprule
%     Data & SNR & Strong & & \multicolumn{2}{c}{E2E (Ours)}\\
%         &  &  Accuracy & Accuracy & mAP \\
%         \midrule
%         Clean & $\infty$ & 96.21 & 97.03 & 98.28 \\
%         \hline
%         $\text{Weak}_{\text{SNR}}$-3s & \multirow{3}{*}{0} &  81.65 & \textbf{93.69} & 95.38\\
%         $\text{Weak}_{\text{SNR}}$-5s & & 73.06 & \textbf{90.78} & 92.01 \\
%         $\text{Weak}_{\text{SNR}}$-7s & & \textbf{67.29} & 62.44 & 10.10 \\
%         \hline
%         $\text{Weak}_{\text{SNR}}$-3s & \multirow{3}{*}{5} & 89.37 & \textbf{95.25} & 96.70 \\
%         $\text{Weak}_{\text{SNR}}$-5s & & 84.48 & \textbf{94.06} & 96.24 \\
%         $\text{Weak}_{\text{SNR}}$-7s & & 72.59 & \textbf{93.45} & 94.87 \\
%         \hline        
%         $\text{Weak}_{\text{SNR}}$-3s & \multirow{3}{*}{10} &  92.51  & \textbf{96.36} & 97.77\\
%         $\text{Weak}_{\text{SNR}}$-5s & & 87.37 & \textbf{95.29} & 96.84 \\

%         $\text{Weak}_{\text{SNR}}$-7s & & 78.17 & \textbf{94.56} & 96.40 \\
%         \bottomrule
%     \end{tabular}
%     \caption{Experiments using keywords being overlapped with noise given a specified SNR. All E2E approaches use 1s random crops.}
%     \label{tab:noisy_experiments}
% \end{table}

The results are displayed in \Cref{tab:noisy_experiments}.
As we can see, additional noise does indeed decrease the performance compared to clean-target target as seen in \Cref{tab:weakly_supervised_baseline}.
A trend can be seen that with the increase of noise (lower SNR), the performance overall consistently decreases.
Further, the performance also uniformly drops when using longer input samples i.e., from 3 s to 7 s.
Finally, we observe that both FA and oracle baselines underperform against our proposed weakly supervised E2E approaches for most training experiments, except for $\text{Weak}_{0}$-7s.
The $\text{Weak}_{0}$-7s result displays the limits of temporally weakly supervised training, where the additional noise on top of the unknown position of the target label has a significant effect on the model, leading to an accuracy of 62.44\%.

However, overall these results using noisy training data are surprising, since the performance decrease is marginal considering that the model had no access to any clean data.
For example, training with 3s long samples with a SNR of 10 db results in a drop of 0.7\% absolute from 97.03\% to 96.36\% and training with a SNR of 0 db accuracy further drops to 93.69\%.
% Further, training with a SNR of 0 dB, leads to an accuracy drop by 3.3\% from 97.03 to 93.69 with 3 s long training samples.
Within our experimented setup, we can conclude that if a target keyword is at least present in more than 15\% ($\frac{1}{7}$) of an utterance with an SNR higher than 0 db, a weakly supervised training approach can successfully outperform our proposed strongly-supervised oracle baseline.
More generally, our results indicate that temporally weak labeled datasets are not necessarily inferior to their strongly supervised counterparts.

\subsection{Ablation study}

Due to the surprising results in \Cref{ssec:train_test_crops} and \Cref{ssec:additive_noise}, we aim to further study the core reasons why temporally weakly-supervised learning works so well.
For this experiment, we create another training set $\text{Weak}_{\text{pos}}-t$s, where only the target keywords (positive samples) are spliced with additional noise (identical to $\text{Weak}-t$s in \Cref{tab:train_test_crops_mismatch}), while non-target keywords (negative samples) are taken from the clean training set.

\begin{table}[htbp]
    \centering
    \begin{tabular}{l|rr}
    \toprule
     Data &  Accuracy & mAP \\
        \midrule
        Clean & 97.03 & 98.28 \\
        \hline
         $\text{Weak}_{\text{pos}}$-3s & 92.08 & 95.21\\
         $\text{Weak}_{\text{pos}}$-5s & 87.80 & 91.40 \\
         $\text{Weak}_{\text{pos}}$-7s & 86.20 & 90.50 \\
        \bottomrule
    \end{tabular}
    \caption{Ablation results which use weak labeling for the target keyword samples, while non-target samples are the original, clean samples. Experiments use 1s crops during training.}
    \label{tab:ablation1_negative_clean}
\end{table}

The results of this ablation study are seen in \Cref{tab:ablation1_negative_clean}.
In our point of view, these results represent our expected behavior regarding weak labeling: longer weakly labeled input samples lead to consistent performance degradation.
Nonetheless, these results are inconsistent with our previous observations in \Cref{tab:train_test_crops_mismatch} and \Cref{tab:noisy_experiments}.

We believe that our findings can be explained as a form of label smoothing~\cite{Szegedy_label_smoothing,NEURIPS2019_f1748d6b}.
In some of our experiments randomly cropping a weakly labeled sample containing a target-keyword has a high likelihood of containing a false positive.
The most striking example of this behavior can be found in our Weak-7s experiments (see \Cref{tab:train_test_crops_mismatch}).
In those experiments, we have a $\frac{6}{7} \approx 85\%$ chance of retrieving a false-positive sample (noise) within the target.
Recall that in the experiments in \Cref{tab:weakly_supervised_baseline}, the ratio between a target keyword and an unknown keyword is 1:20.
This means that if the model is presented a false-positive sample, the overall probability that this sample is correct according to the target/non-target ratio of the dataset is also 1:20, leading to a rejection of the false-positive sample by the model.
However, this behavior only seems to occur if non-target samples are provided capable of helping the model detect those false-positives, i.e., noisy samples from Audioset found within our non-target keywords.
When one removes the access to those negative samples, the performance significantly drops as seen in \Cref{tab:ablation1_negative_clean}.

\section{Discussion}
\label{sec:discussion}

This work's results encourage training audio classification models in a weakly-supervised fashion.
We strongly believe that our results also transfer to similar tasks within the audio domain, such as acoustic event detection~\cite{mesaros2010acoustic} and voice activity detection~\cite{chen2020voice,dinkel2021voice,xu2021lightweight} and acoustic environment classification~\cite{piczak2015esc}.
We provided empirical evidence that temporally weakly supervised learning can be seen as a form of label smoothing.
However, label smoothing adjusts the target label for a clean sample, while weakly supervised learning allows collecting data without requiring pin-point accuracy, potentially increasing the available data.
It is important to note that the FA baseline is not entirely comparable to our proposed approach, since its label acquisition process is automatic, while our proposed approach requires manual human annotation.
Future work might aim to merge these two approaches, by i.e., including misaligned samples into the temporally weak training process.

\section{Conclusions}
\label{sec:conclusion}

This work studies the effects of temporally weak supervision specifically on KWS models.
Our experiments compare a standard HMM baseline approach to strongly and weakly supervised E2E approaches.
Multiple artificial datasets are constructed to investigate the effects of temporally weak labeling.
One of our key results demonstrates that weak labeling is preferable to strongly supervised approaches in presence of strong noise.
Surprisingly, our results suggest that temporally weak supervised training can improve KWS performance. 
As a conclusion, we suggest the following training techniques for KWS:
\begin{itemize}
    \item Additional silence or noise within a training sample has little impact on an E2E model's performance and thus can be kept in order to enlarge the training dataset.
    \item If the target keyword length is provided, random cropping the input sequence with that target keyword length provides a generally better KWS performance.
    \item Even small CNN models are capable of localizing and identifying a target keyword if that target keyword is present in at least 15\% of the input audio clip duration.
    \item In presence of strong noise, where force-alignments are unreliable, one can consider manually labeling data in a coarse fashion and use weakly supervised training to significantly enhance performance.
\end{itemize}

%\section{Acknowledgements}

\bibliographystyle{IEEEtran}

\bibliography{bib}

% \begin{thebibliography}{9}
% \bibitem[1]{Davis80-COP}
%   S.\ B.\ Davis and P.\ Mermelstein,
%   ``Comparison of parametric representation for monosyllabic word recognition in continuously spoken sentences,''
%   \textit{IEEE Transactions on Acoustics, Speech and Signal Processing}, vol.~28, no.~4, pp.~357--366, 1980.
% \bibitem[2]{Rabiner89-ATO}
%   L.\ R.\ Rabiner,
%   ``A tutorial on hidden Markov models and selected applications in speech recognition,''
%   \textit{Proceedings of the IEEE}, vol.~77, no.~2, pp.~257-286, 1989.
% \bibitem[3]{Hastie09-TEO}
%   T.\ Hastie, R.\ Tibshirani, and J.\ Friedman,
%   \textit{The Elements of Statistical Learning -- Data Mining, Inference, and Prediction}.
%   New York: Springer, 2009.
% \bibitem[4]{YourName17-XXX}
%   F.\ Lastname1, F.\ Lastname2, and F.\ Lastname3,
%   ``Title of your INTERSPEECH 2022 publication,''
%   in \textit{Interspeech 2022 -- 23\textsuperscript{rd} Annual Conference of the International Speech Communication Association, September 18-22, Incheon, Korea, Proceedings, Proceedings}, 2022, pp.~100--104.
% \end{thebibliography}

\end{document}